# Maximum sustainable yields from a spatially-explicit harvest model

by
Nao Takashina[1*,] and Akihiko Mougi[2]

1. *Department of Biology, Faculty of Sciences,
   Kyushu University, 6-10-1, Hakozaki, Fukuoka, 812-8581, Japan,
   nao.takashina@gmail.com*

2. *Department of Biological Science, Faculty of Life and Environmental Science,
   Shimane University, Nishikawatsu-cho 1060 Matsue 690-8504 Japan,
   amougi@gmail.com*

**Abstract**

Spatial heterogeneity plays an important role in complex ecosystem dynamics, and therefore is also an important consideration in sustainable resource management. However, little is known about how spatial effects can influence management targets derived from a non-spatial harvest model. Here, we extended the Schaefer model, a conventional non-spatial harvest model that is widely used in resource management, to a spatially-explicit harvest model by integrating environmental heterogeneities, as well as species exchange between patches. By comparing the maximum sustainable yields (MSY), one of the central management targets in resource management, obtained from the spatially extended model with that of the conventional model, we examined the effect of spatial heterogeneity. When spatial heterogeneity exists, we found that the Schaefer model tends to overestimate the MSY, implying potential for causing overharvesting. In addition, by assuming a well-mixed population in the heterogeneous environment, we showed analytically that the Schaefer model always overestimate the MSY, regardless of the number of patches existing. The degree of overestimation becomes significant when spatial heterogeneity is marked. Collectively, these results highlight the importance of integrating the spatial structure to conduct sustainable resource management.

## 1. Introduction

Diverse ranges of environments are characterized by spatial heterogeneity. Ecologists recognize that this heterogeneity plays a critical role in the complex dynamics of ecosystem (Hanski, 1998; Levin, 1992), and also it is practically an important consideration in ecosystem management (Plotkin and Muller-Landau, 2002). The use of spatially explicit approaches to the ecosystem management are increasing rapidly in response to the recent trends to involve reserves in terrestrial as well as marine ecosystem management (Baskett and Weitz, 2007; Lundberg and Jonzén, 1999; Neubert, 2003; Sanchirico and Wilen, 1999; Takashina et al., 2012; White and Costello, 2011; White et al., 2010; Williams and Hastings, 2013). On the other hand, in many management exercises, including fisheries management (Clark, 1990; Walters et al., 2005) and terrestrial wildlife hunting (Ling and Milner-Gulland, 2008; Robinson and Redford, 1991), managers traditionally use the concept of maximum sustainable yield (MSY) without consideration of the spatial structure. This is likely because most harvesting theories, in which MSY has played a major role in sustainable resource uses, originated in commercial fisheries science (Gordon, 1954; Schaefer, 1954) where spatial heterogeneity was not considered until recently (Ling and Milner-Gulland, 2008). Therefore, this leaves us to question how integrating the spatial structure affects the management goals in the harvesting model.

---

[*] Corresponding author

Ling and Milner-Gulland (2008) used a static spatial harvesting model but also considered the effects of traveling costs, showing that MSY can be overestimated when these costs are not taken into account. Ying et al. (2011) discussed the risks of ignoring spatial structure in a 10-year simulation of fisheries management, showing that such an omission resulted in a high probability of fishing stocks off the coast of China being over exploited and/or suffering localized depletions. Both papers highlight the importance of explicitly considering spatial structure in mitigating the risks of overestimation or overexploitation with a specific setting in mind. Křivan and Jana (2015) discussed the effect of the dispersal on harvesting with the no-take marine reserve where the two regions (the fishing ground and reserve) are characterized by the proportional size of the concerned area. They showed numerically that the dispersal of the species could lead to the decline of the population abundance as well as the MSY.

In this study, we developed a general spatially-explicit model which is naturally extended by the conventional (non-spatial) harvest model, and therefore we can apply it to various resource managements. One of the conventional models used widely in resource management is the Schaefer model (Clark, 1990; Schaefer, 1954). In addition, this model is often used as a basis for more complex ecosystem models (Neubert, 2003). In light of this, it may be rational to extend the Schaefer model to include a spatial structure as the first step towards the spatial extension of harvest models.

In this paper, we examine the spatial effect on the MSY of a harvesting model by extending the Schaefer model to a spatially generalized model. We show that when spatial structure is not considered, this omission leads to an overestimation in MSY, implying potential for causing overharvesting by providing larger amount of harvestable population. We also discuss the conditions in which the overestimation becomes significant, and a way to apply our model to an actual management to predict degree of the overestimation.

## 2. Methods
*2.1 MSY in the Schaefer model*
One of the most basic harvest models is the Schaefer model, which can be described as:
$$\frac{dx}{dt} = rx\left(1 - \frac{x}{K}\right) - ex, \qquad (1)$$

where $x$ is population abundance, $r$ the per capita growth rate (per unit time), $K$ is the carrying capacity of the environment and $e$ is the harvest rate (per unit time). Using this equation, MSY is calculated to be equal to $rK/4$ and thus, when MSY is reached, population abundance is equal to $K/2$ (Gordon, 1954; Schaefer, 1954).

*2.2 The spatially explicit harvest model*
In this study, we considered a simple spatial generalization of the Schaefer model, hereafter referred to as the spatially explicit harvest model (SEH). One of the simplest ways to spatially extend a non-spatial model is to divide the area being considered into two patches with the area fractions $\alpha$ and $1-\alpha$, and each of these patches is assumed to have different habitat qualities $K_1$ and $K_2$ (per unit area). It is worth stressing that the carrying capacity $K$ in Eq. (1) and these habitat qualities are not the same quantities due to the difference in their units. The carrying capacities in the SHE model are then the product of the area fraction and the habitat quality in the patch (Fig. 1), and therefore $K = \alpha K_1 + (1-\alpha) K_2$. The two patches are interconnected through the exchange of individuals from the two populations, an event that is represented by the exchange rate $m$, defined for each time period and each patch. Therefore, the actual exchange rate between populations is proportional to the area of other patch and the population abundance $x_i$ ($i = 1, 2$) in the focal patch. We add the exchange terms to the Schaefer model (Eq. 1) to obtain the two-patch SEH model:

$$\frac{dx_1}{dt} = rx_1\left(1 - \frac{x_1}{\alpha K_1}\right) - e_1 x_1 + m(\alpha x_2 - (1-\alpha) x_1), \qquad (2a)$$

$$\frac{dx_2}{dt} = rx_2\left(1 - \frac{x_2}{(1-\alpha) K_2}\right) - e_2 x_2 + m((1-\alpha) x_1 - \alpha x_2). \qquad (2b)$$

The subdivision of the area does not change $r$ and managers can take different harvest rates $e_i$ for each patch.

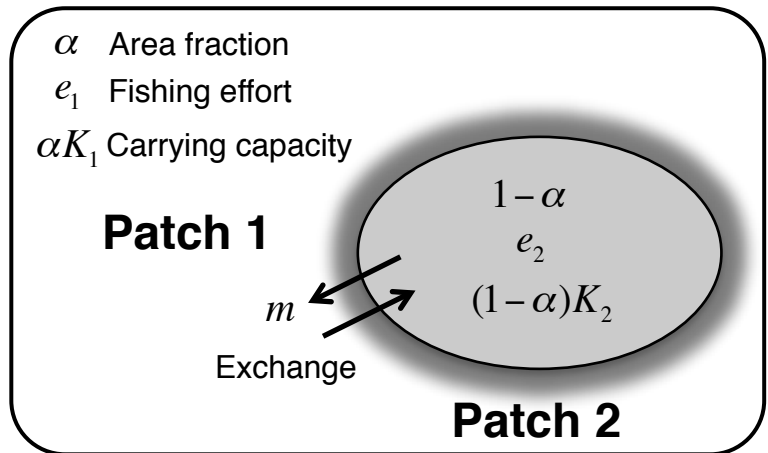

**Figure 1** Schematic description of the spatial-integrated Schaefer model. Environmental heterogeneities create two different patches in the concerned area. Two patches have different habitat qualities $K_i$ ($i$=1, 2) and fractions of the area $\alpha$, $1-\alpha$ and exchange of species connect with patches at rate *m*.

*2.3 The Schaefer model versus the SEH model*
To examine the effects of spatial differences on MSY we compared the MSYs calculated by both the Schaefer and SEH models. In the SEH model, the conventional MSY becomes $r(\alpha K_1 + (1-\alpha)K_2)/4$, noting that the *K* in Equation 1 has been replaced by the total carrying capacity of the whole area. For simplicity, *r* was set at unity, but it does not change the ratio between the conventional MSY and the MSY in the SHE model because it does not appear in the ratio.

For the SEH model, it was possible to calculate two different MSY values depending on which management regime was applied. In the first regime (uMSY) harvest rates were assumed to be uniform for both patches (i.e., $e_1 = e_2$) whilst for the second regime (gMSY) harvest rates were altered in both patches with a view to reaching a global MSY that was defined as the MSY in the whole area.

In the following section, we examine the effect of space on MSY in cases where the two patches are isolated, connected through an intermediate exchange rate or well-mixed by a high exchange rate. For more general situations, we also considered an *n*-patch generalization of the SHE model for broader applications. We did not examine the population abundance at the MSY values, because the MSY in the Schaefer model is proportional to the population abundance at the MSY value and one may infer the spatial effect to the population abundance at the MSY.

## 3. Results
### *3.1 The two-patch SHE model*
*3.1.1. Isolated patches*
When the two sections are isolated from each other (i.e., thus *m* = 0 in Eq. 2a and 2b), uMSY and gMSY are simply the sum of the MSY values calculated independently for each section. In this case, it is then clear that uMSY is equal to gMSY, and also that they are both equal to the conventional MSY. Thus, when the two sections are isolated, there are no spatial effects on the MSY in either management regime.

*3.1.2a Interconnection via an intermediate exchange rate*
When the two sections are interconnected by an intermediate exchange rate (i.e., *m* = 1 or 10), uMSY is always smaller than the conventional MSY. The decrease in uMSY, relative to the conventional MSY, becomes significant when there is an intermediate size ratio between the two patches (Fig. 2a). Addionally, it also becomes larger when the difference in the habitat qualities of the two patches is large or the exchange rate is relatively high. These results suggest that an overestimation in the MSY is likely to occur when the target population moves easily between habitats with high environmental heterogeneity. In this regime, using the conventional MSY could result in an overestimation of up to 2.85 times (Fig. 2a; triangles). The gMSY shows a similar pattern when the exchange rate and habitat quality ratio are both relatively high (i.e., *m* = 10 and $K_2/K_1 = 10$; Fig. 2a). Here, using the conventional MSY could result in an overestimation that was approximately 2.63-times the true MSY (Fig. 2b; triangles). However, unlike the uMSY, gMSY does not show significant declines relative to the conventional MSY when the exchange rate and c habitat quality ratio are relatively small (i.e., $m = 1$ and $K_2 / K_1 = 2$; Fig. 2b).

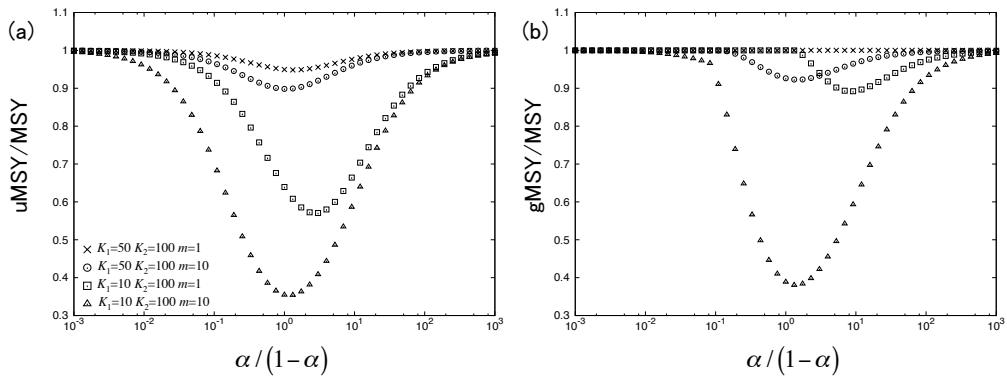

**Figure 2** uMSY (a) and gMSY (b) values relative to the conventional MSY.

*3.1.2b Harvest rates*
The harvest rates of each patch change according to the size ratio of the two patches (i.e., $\alpha/(1-\alpha)$) in the gMSY regime (Fig. 3). As the fraction of space taken by Patch 1 increased, the harvest rates in both Patch 1 and 2 approach the 0.5 and 0 MSY values obtained from the Schaefer model, respectively. This occurs regardless of what the habitat quality and exchange rate values are (Fig. 3). Conversely, when the ratio of Patch 1 decreased, its harvest rate becomes very large. It also tended to diverge, especially when there was a high exchange rate and a relatively large difference in the habitat qualities of the two patches (Fig. 3a; triangles; maximum harvest rate was set at 20 to prevent the harvest rate diverging). Meanwhile, in Patch 2, the harvest rate approached the MSY value obtained from the Schaefer model.

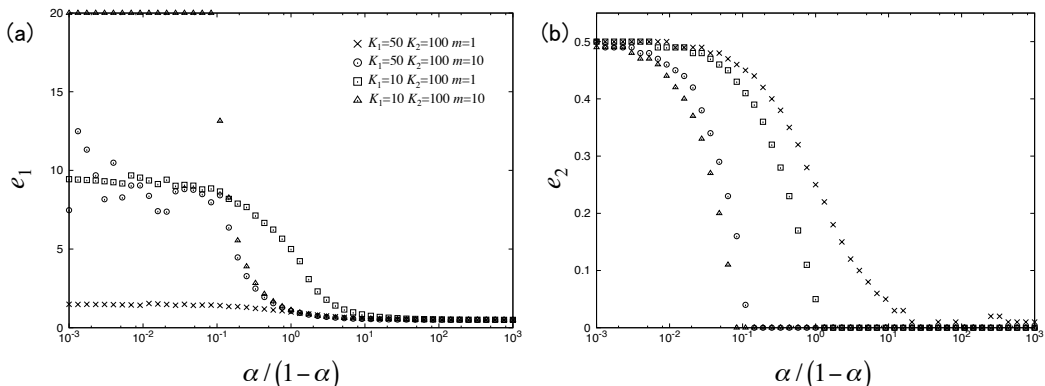

**Figure 3** Harvest rates in (a) patch 1 and (b) patch 2 in the regime of gMSY management.

*3.1.3 A well-mixed population (through a high exchange rate)*

When the population exchange rate between the two sections is sufficiently high (i.e., $m \gg 1$), we can approximate the dynamics of the two populations (Eqs. 2a and 2b) into one population dynamic model using the aggregation method: this method assumes that $m$ has a negligible effect on the overall dynamics of the entire population's abundance (i.e., the macro variable, $X = x_1 + x_2$) at the time scale $\tau$ (Auger et al., 2008; Iwasa and Andreasen, 1987). Thus, we can formulate the simple equation:

$$\frac{dX}{d\tau} = rX\left(1 - \frac{X}{K'}\right) - e'X, \tag{3}$$

where $K' = K_1 K_2 / (\alpha K_2 + (1-\alpha) K_1)$ and $e' = e_1 \alpha + e_2(1-\alpha)$. In this equation we do not impose any requirements on $e_i$, (as assumed in the uMSY regime). Using the MSY value obtained from the aggregation model (MSY'), we can then calculate the ratio of MSY' to the conventional MSY using the following equation:

$$\begin{aligned}
\frac{MSY'}{MSY} &= \frac{K_1 K_2}{(\alpha K_1 + (1-\alpha) K_2)(\alpha K_2 + (1-\alpha) K_1)} \\
&= \frac{1}{\left(\alpha + (1-\alpha)\frac{K_2}{K_1}\right)\left(\alpha + (1-\alpha)\frac{K_1}{K_2}\right)} \\
&= \frac{1}{\alpha^2 + (1-\alpha)^2 + \alpha(1-\alpha)\left(\frac{K_1^2 + K_2^2}{K_1 K_2}\right)} \\
&= \frac{1}{2\alpha(\alpha-1)\left(1 - \frac{K_1^2 + K_2^2}{2 K_1 K_2}\right) + 1} \\
&= \frac{1}{\alpha(1-\alpha)\frac{(K_1 - K_2)^2}{K_1 K_2} + 1} \\
&\leq 1.
\end{aligned} \tag{4}$$

This calculation clearly shows that MSY' does not exceed the conventional MSY. When the habitat qualities in both patches are equal (i.e., $K_1 = K_2$) there is no spatial effect, regardless of patch sizes. By calculating the right-hand side of Equation 4, we found that the decline in MSY' relative to the conventional MSY becomes significant as the ratio of $K_1 / K_2$ becomes larger or smaller. This effect is magnified if the patch with the larger habitat quality then also expands in size. We confirmed that when $m$ is high (i.e., $m = 10^3$; Fig. A1 in Appendix) the equality of Equation 4 was a good fit for the values obtained from the simulations.

*3.2 The n-patch SEH model*

The two-patch SEH model can be extended to a more general *n*-patch model in which population exchanges between patches occurs at a rate proportional to the area of another patch and the population abundance in the focal patch. Thus, given the fraction of the specified area $\alpha_i$, a habitat quality of $K_i$, and a harvest rate $e_i$ in the $i^{th}$ patch, the population abundance dynamics in that patch $x_i$ can be described as:

$$\frac{dx_i}{dt} = rx_i\left(1 - \frac{x_i}{\alpha_i K_i}\right) - e_i x_i + m\left(\alpha_i \sum_{j \neq i}^{N} x_j - (1-\alpha_i)x_i\right), \quad (1 \leq i \leq n), \tag{5}$$

where the dimensions of the model parameters are the same as in the two-patch SEH model (Eqs. 2a and 2b) and $\alpha_1 + \cdots + \alpha_n = 1$. A similar aggregation method can also be applied when *m* is sufficiently high (i.e., $m \gg 1$), giving us:

$$\frac{d\hat{X}}{d\tau} = r\hat{X}\left(1 - \frac{\hat{X}}{\hat{K}}\right) - \hat{e}\hat{X}, \tag{6}$$

where $\hat{K} = \sum_{i}^{n} \alpha_i K_1 \cdots K_{i-1} K_{i+1} \cdots K_n \Big/ \prod_{i=1}^{n} K_i$ and $\hat{e} = \sum_{i}^{n} e_i \alpha_i$. For all values of *i*, we did not impose any restrictions on $e_i$ and assumed that $K_i > 0$. The MSYs obtained from the *n*-patch SEH model and Schaefer model were denoted as $MSY'_n$ and $MSY_n$, respectively. As before, we also calculated the $MSY'_n / MSY_n$ ratio using the following equation:

$$\frac{MSY'_n}{MSY_n} = \frac{\prod_{i}^{n} K_i}{\sum_{i,j}^{n} \alpha_i K_i \alpha_j K_1 \cdots K_{j-1} K_{j+1} \cdots K_n}$$

$$= \frac{1}{\sum_{i}^{n}\sum_{j\neq i}^{n} \alpha_i K_i \alpha_j \frac{1}{K_j} + \sum_{i}^{n} \alpha_i^2}$$

$$= \frac{1}{\sum_{i}^{n-1}\sum_{j>i}^{n} \alpha_i \alpha_j \left(\frac{K_i}{K_j} + \frac{K_j}{K_i}\right) + \sum_{i}^{n} \alpha_i^2}$$

(a) $\leq \dfrac{1}{2\sum_{i}^{n-1}\sum_{j>i}^{n} \alpha_i \alpha_j + \sum_{i}^{n} \alpha_i^2}$ \hfill (7)

(b) $= \dfrac{1}{\sum_{i,j}^{n} \alpha_i \alpha_j}$

$= \dfrac{1}{\sum_{i}^{n} \alpha_i}$

$= 1.$

To obtain the inequality (a), we use the fact $(K_i^2 + K_j^2)/K_i K_j \geq 2$ with a requirement $K_i, K_j \neq 0$ for all $i$ and $j$ and an equality occurs if $K_i = K_j$. We can conclude that if condition (a) is satisfied ($K_i = K_j$ for all $i$ and $j$) the distribution of the areal size, $\alpha$, does not matter. Eq. (7) shows that $MSY'_n$ never exceeds $MSY_n$, regardless of the number of patches, and this supports the general applicability of the two-patch SHE model.

## 4. Discussion

In this paper, we extended the Schaefer model (Gordon, 1954; Schaefer, 1954) to a general spatial model that allowed us to examine the spatial effect on MSY. Our results showed that when the harvest model lacked spatial structure, it tended to overestimate MSY, regardless of whether plausible management regimes were in place. In the most extreme case, the Schaefer model overestimated MSY by about 2.85. Overestimation is more likely to occur when spatial heterogeneity is greater, and the manager applies a homogeneous effort level in the whole area (uMSY). Furthermore, when the populations in the area of concern are well-mixed through rapid exchanges, we showed analytically that overestimating MSY always occurs, regardless of the number of patches existing.

Looking explicitly at spatial structure, the management regimes that may plausibly be available are uMSY and gMSY: the specific choice depends on the resolution of the spatial information available. If the data are of a sufficiently fine scale to detect environmental heterogeneity, managers can choose an appropriate management regime that avoids the overestimation of MSY. However, it is worth noting that gathering such fine-scale spatial information and then applying fine-tuned management responses are much more expensive than relying on coarse resolution data and uniformed-effort management (Naidoo et al., 2006; Richardson et al., 2006). Therefore, managers should always take such costs into consideration when making decisions.

Our model may include a property similar to that of the optimal harvesting model in terms of the two-patch, source-sink dynamics (Lundberg and Jonzén, 1999). Our two-patch model displays source-sink properties when the size of one patch becomes much bigger because, in such a case, emigration to the other patch becomes almost zero as immigration into the larger patch immediately experiences a negative growth rate. This is because of changes to the carrying capacity (fraction of the area × habitat quality). They concluded that the optimal harvest strategy occurs when either (a) the sink is harvested at the MSY level or (b) the sink is harvested at a maximum rate and the source is harvested at its MSY level. Our model shows a similar trend to the first proposal because as $\alpha/1-\alpha$ becomes very small, the richer patch becomes the source (Fig. 3). Conversely, regardless of parameter values, all the curves show a trend in line with the second strategy proposed when $\alpha/1-\alpha$ becomes very large, with the poorer patch becoming the source. We observed this tendency towards (b) starts when $\alpha/1-\alpha$ reaches (and exceeds) 1 and the exchange rate is high (i.e., $m$=10). This suggests that a larger exchange rate promotes source-sink dynamics in cases where the poorer patch has become the source.

We used an aggregation method to obtain analytical relationship between the MSY of spatially structured model and that of the conventional model. When $m$ is $10^2$, the simulated values of uMSY/MSY show a good fit to the analytical values, but the simulated values with $m=10$ also show a relatively good fit (Fig. A1 in the Appendix). An exchange rate of $m \geq 10^2$ may not be unusual for species with highly mobility, a large home range, and/or species that use distinct foraging and refuge areas. Species that have long periods between breeding events and/or late-maturing also tend to have larger exchange rates because the value in exchange rate tends to be larger with the time scale of the reproduction event. In marine ecosystem, for example, species of large body size are likely to exploit resources over larger areas (Kramer and Chapman, 1999; Lowe and Bray, 2006; Sale et al., 2005), and moderately and highly mobile species are generally long-lived, slow-growing and late-maturing such as cod, snappers, or groupers (Gruss et al., 2011; Polunin, 2002; Sale et al., 2005).

For practical applications, an easy way to make use of the relationship described in Equation 7 is to subdivide a given management area into *n* equal-sized patches, namely $\alpha_i = 1/n$ for all *i*. By substituting this into the right hand-side of the equation presented in the first line of Equation 7 and performing some basic algebra, the following equation can be obtained: $MSY'_n/MSY = 1 \bigg/ \left( \dfrac{1}{n^2} \left( \sum\limits_{i}^{n-1} \sum\limits_{j>i}^{n} \left( \dfrac{K_i^2 + K_j^2}{K_i K_j} \right) \right) + n \right)$. Having completed this step, the only unknown parameters are the habitat qualities, and therefore one can estimate the decline from the MSY value by measuring the habitat qualities in each patch. Managers can choose an arbitrary number of subdivisions $n$: their selection may depend on the existing data available or the technological limitations of measuring habitat quality. Note, however, that the approximated value $MSY'_n/MSY$ may become more reliable as $n$ increases because the exchange rate $m$ increases with the number of subdivisions.

The SHE model explored in this paper is one of the simplest extensions of the conventional harvest model. However, despite its simplicity, it can provide many important predictions. Our outputs strongly support the importance of incorporating spatial structure into harvest model and provide more reliable population dynamics of the harvested population. In natural systems, the habitat heterogeneity is widely observed, and it may tend to increase with a size of the focal area (spatial scale of the management region). Spatial resolution of data also affects available management regime, suggesting importance of decision-making of the spatial unit scale in the environmental assessment for each management region. Applying spatial explicit harvest model helps avoiding overharvesting by providing overestimated MSY, and it will lead to the sustainable resource uses.

## Acknowledgements

This work was supported by a Grant-in-Aid for Japan Society for the Promotion of Science (JSPS) Fellows granted to NT. We thank L. Barnett, M.L. Baskett, Y. Tachiki, C. Chou, and Y. Wang for their thoughtful comments.

**Appendix Figure**

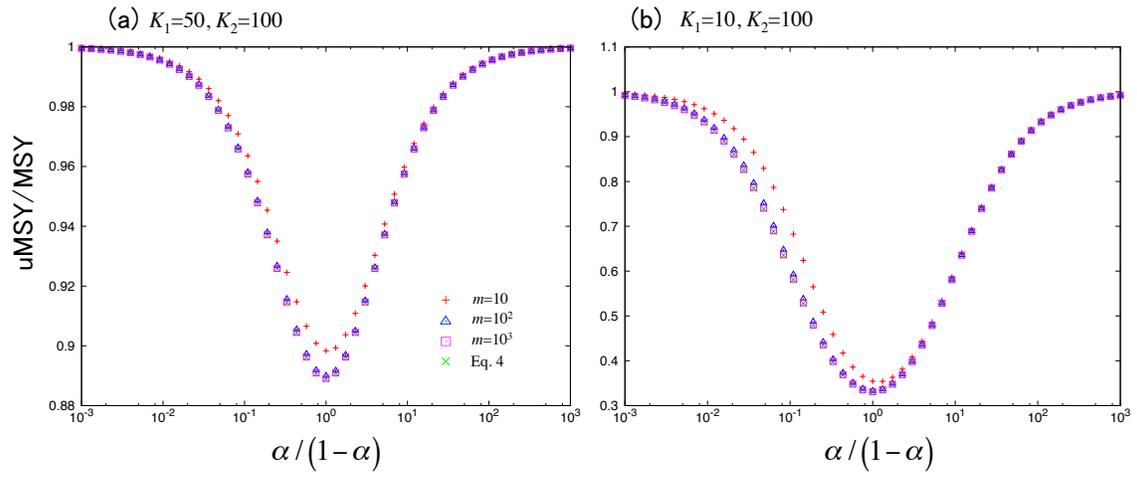

**Figure A1** uMSY values relative to the conventional MSY for various exchange rates ($m = 10, 10^2, 10^3$) and analytic value (Eq. 4). (a) $K_1 = 10, K_2 = 100$, (b) $K_1 = 50, K_2 = 100$.